\documentclass[a4paper]{article}
\usepackage[T1]{fontenc}
\usepackage{authblk}
\usepackage{graphicx}

\title{The discrimination capabilities of Micromegas detectors at low energy}
\author[1]{F.J.~Iguaz}
\author[2]{T.~Dafni}
\author[1]{E.~Ferrer-Ribas}
\author[1]{J.~Gal\'an}
\author[2]{J.A.~Garc\'ia}
\author[3]{A.~Gardikiotis}
\author[1]{I.~Giomataris}
\author[2]{I.G.~Irastorza}
\author[1]{J.P.~Mols}
\author[1]{T.~Papaevangelou}
\author[2]{A.~Rodr\'iguez}
\author[2]{A.~Tom\'as}
\author[3,4,5]{T.~Vafeiadis}
\author[6]{C.~Yildiz}

\affil[1]{IRFU, Centre d'\'Etudes Nucl\'eaires de Saclay (CEA-Saclay), Gif-sur-Yvette, France}
\affil[2]{Laboratorio de F\'isica Nuclear y Astropart\'iculas,
Universidad de Zaragoza, Spain}
\affil[3]{University of Patras, Greece}
\affil[4]{European Organization for Nuclear Research (CERN), Gen\`eve, Switzerland}
\affil[5]{Aristotle University of Thessaloniki, Greece}
\affil[6]{Do\u{g}u\c{s} University, Istanbul, Turkey}

\begin{document}

\maketitle

\begin{abstract}
The latest generation of Micromegas detectors show a good energy resolution, spatial resolution and low threshold, which make them idoneous in low energy applications. Two micromegas detectors have been built for dark matter experiments: CAST, which uses a dipole magnet to convert axion into detectable x-ray photons, and MIMAC, which aims to reconstruct the tracks of low energy nuclear recoils in a mixture of CF$_4$ and CHF$_3$. These readouts have been respectively built with the microbulk and bulk techniques, which show different gain, electron transmission and energy resolutions. The detectors and the operation conditions will be described in detail as well as their discrimination capabilities for low energy photons will be discussed.
\end{abstract}

\section{Micromegas: the bulk and microbulk technologies}
\label{sec:micromegas}
Micromegas (for MICRO MEsh GAseous Structure) is a parallel-plate detector invented by I. Giomataris in 1995 \cite{Giomataris:1995fq}. It consists of a thin metallic grid (commonly called mesh) and an anode plane, separated by insullated pillars. Both structures define a very little gap (between 20 and 300~$\mu$m), where primary electrons generated in the conversion volume are amplified, applying moderate voltages at the cathode and the mesh. This type of readouts have features like their high granularity, good energy and time resolution, easy construction, little mass and gain stability. The two actual technologies (bulk \cite{Giomataris:2006yg} and microbulk \cite{Adriamonje:2010sa, Iguaz:2011fi}) create all-in-one detectors, in opposition to the first ones for which the mesh was mechanically mounted on top of the pixelised anode.

\medskip
The bulk Micromegas technology allows to yield large area, robust readouts by integrating a commercial woven wire mesh together with the anode carrying the strips or pixels. Two photo-resistive layers with the right tickness are laminated with the anode printed circuit and the mesh at high temperatures, forming a single object. The supporting mesh pillars are formed illuminating the piece by UV photons and the rest of the material is removed by a chemical bath. The result is a robust, homogeneous and reproducible readout that can be made up to relatively large areas at low cost.

\medskip
The microbulk Micromegas is a more recent development that provides double-clad kapton foils. The mesh is etched out of one of the copper layers of the foil, and the Micromegas gap is created by removing part of the kapton by means of appropriate chemical baths and photolithographic techniques. The amplification gap is more homogeneous and the mesh geometry has better quality than bulk detectors. These features make the readouts have the best energy resolutions among MPGDs, as low as 11\% FWHM at 5.9 keV in argon-based mixtures and 10.5\% FWHM in neon-based ones \cite{Iguaz:2011fi}. Moreover, these readouts have showed very low levels of radiopurity \cite{Cebrian:2011sc}. On the other hand, they are less robust than the bulk and the maximum size of single readouts is $30 \times 30$ cm$^2$, due to the actual equipment.

\section{CAST: an axion experiment}
\label{sec:cast}
The CERN Axion Solar Telescope (CAST) \cite{KZioutas:2005kz, Andriamonje:2010sa2, EArik09} uses a prototype of a superconducting LHC dipole magnet to convert axions into detectable x-ray photons. Axions are pseudoscalar particles that appear in the Peccei-Quinn solution of the strong CP problem and are candidates to Dark Matter. The Sun could produce a big flux of axions via the Primakoff conversion of plasma photons. These particles could then couple to a virtual photon provided by the magnetic transverse field of the CAST magnet (9 Teslas), being transformed into a real photon that carries the energy and momentum of the original axion. These photons can be detected by the four x-ray detectors installed at the ends of the 10 m long magnet, when the magnet points at the Sun. This happens twice a day during the sunset and sunrise for a total time of 3 hours per day.

\subsection{Micromegas detectors in CAST}
\begin{figure}[htb!]
\centering
\includegraphics[height=.30\textwidth]{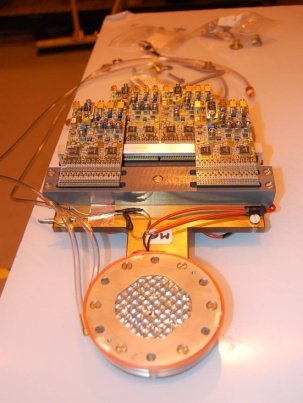}
\hspace{3mm}
\includegraphics[height=.30\textwidth]{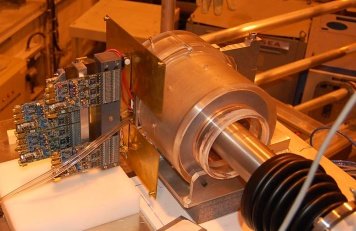}
\hspace{3mm}
\includegraphics[height=.30\textwidth]{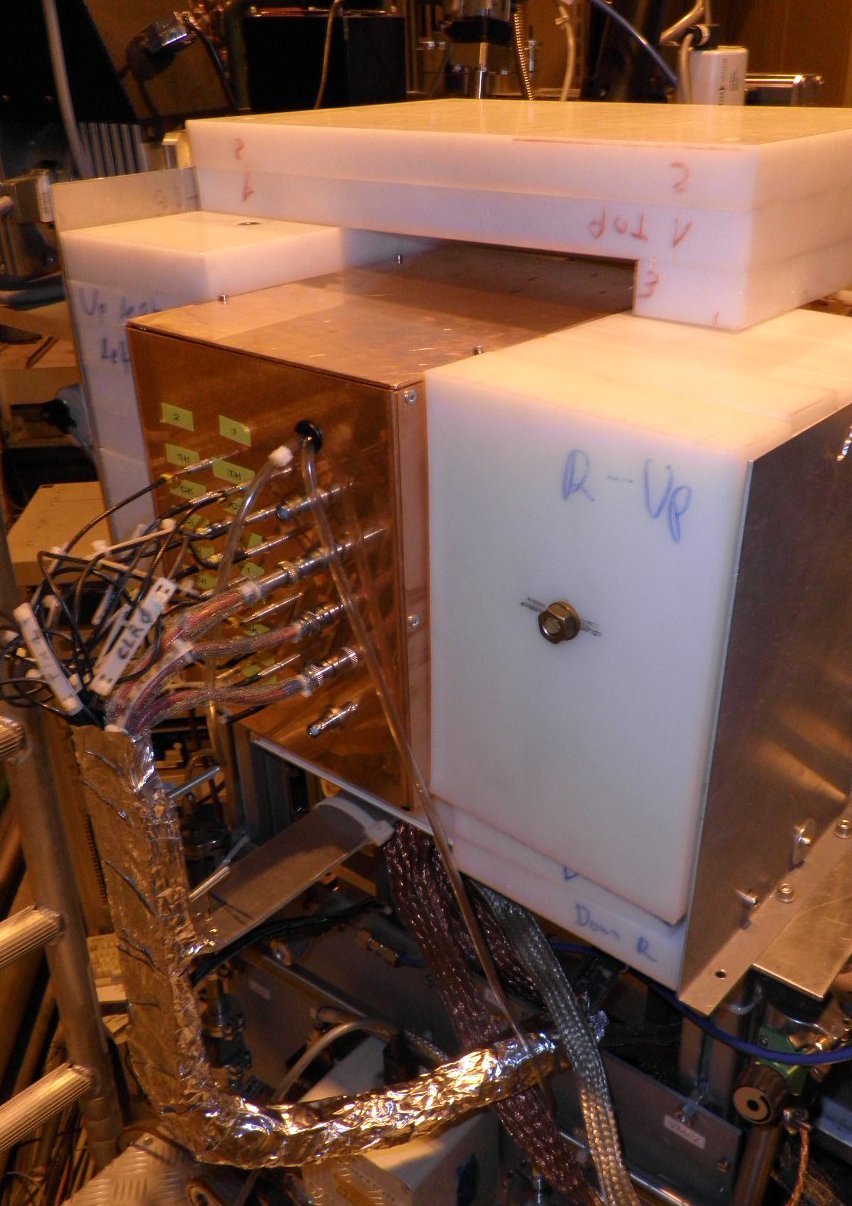}
\caption{Left: The Micromegas detector used in CAST. The active area is situated at the front part and is covered with a stainless steel window and a plexiglass piece. The strips are read by four Gassiplex cards situated at the rear part. Center: The circular lead shielding that surrounds the readout and the stainless steel tube that comes out from the magnet bore. Right: The Faraday cage and the external polyethylene shielding that covers the lead shielding. The electrical signals and the gas tubes are extracted from the copper box via feedthroughs.}
\label{fig:CAST}
\end{figure}

Since 2008, three microbulk Micromegas detectors are being used \cite{SAune:2009sa}, replacing a conventional Micromegas and a TPC. The readouts' anode consists of square pads interconnected in the diagonal directions through vias in two extra back-layers, producing a strip pitch of 550 $\mu$m and 106 strips in each direction. The readout is situated in a TPC, shown in figure \ref{fig:CAST} (left), formed by a cylindrical plexiglas body and a stainless steel strong back that works as drift plate and is also used for the coupling with the magnet bore. The gas used is Ar + 2.3\%iC$_4$H$_{10}$ at 1.44 bar. The chamber is situated inside a 5~mm-thick copper Faraday cage and is shielded by a 25 mm-thick archeological lead shielding and a 15-20 cm of polyethylene, as shown in figure \ref{fig:CAST} (center and right). The whole setup is flushed with nitrogen in order to remove radon.

\medskip
The data acquisition system registers the analogue signal induced in the mesh for each event with a 1 GHz FADC and the integrated charge on each strip using four Gassiplex cards. From both signals, several parameters are extracted like the cluster size, multiplicity and width from the strips data and the risetime, width, amplitude and integral from the mesh pulses. Some of these variables are used in the analysis routines to discriminate x-rays from background events, considering the physics case. An x-ray of less than 10 keV produces a primary ionization localized in a spatial range less than 1 mm. The amplification of this charge gives a narrow pulse with a fixed risetime and a mean strip multiplicity which corresponds to around 5 mm. In contrast, cosmic muons and high energy gammas produce a spatially extended ionization, resulting to broader pulses and higher multiplicities.

\begin{figure}[htb!]
\centering
\includegraphics[height=39mm]{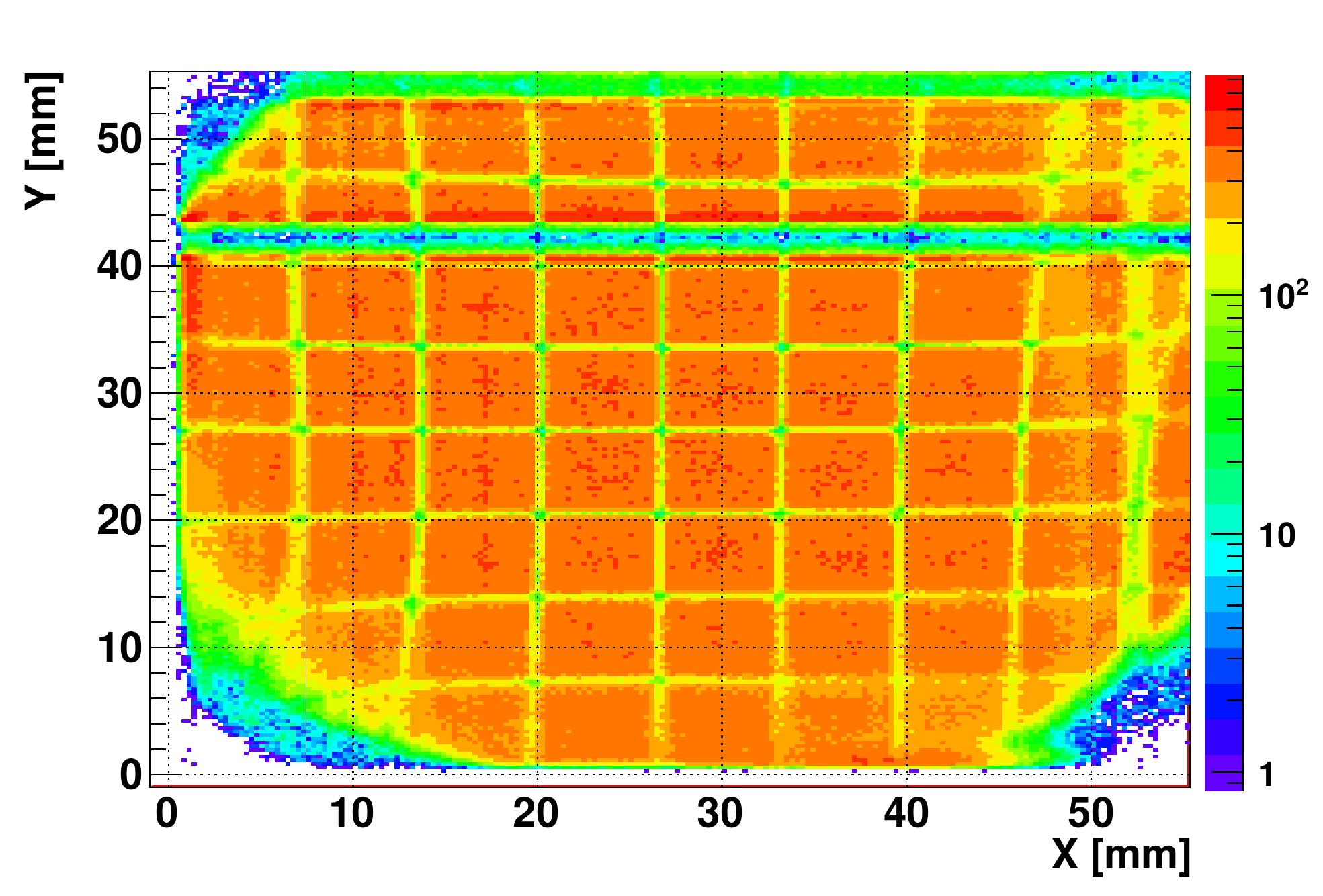}
\includegraphics[height=39mm]{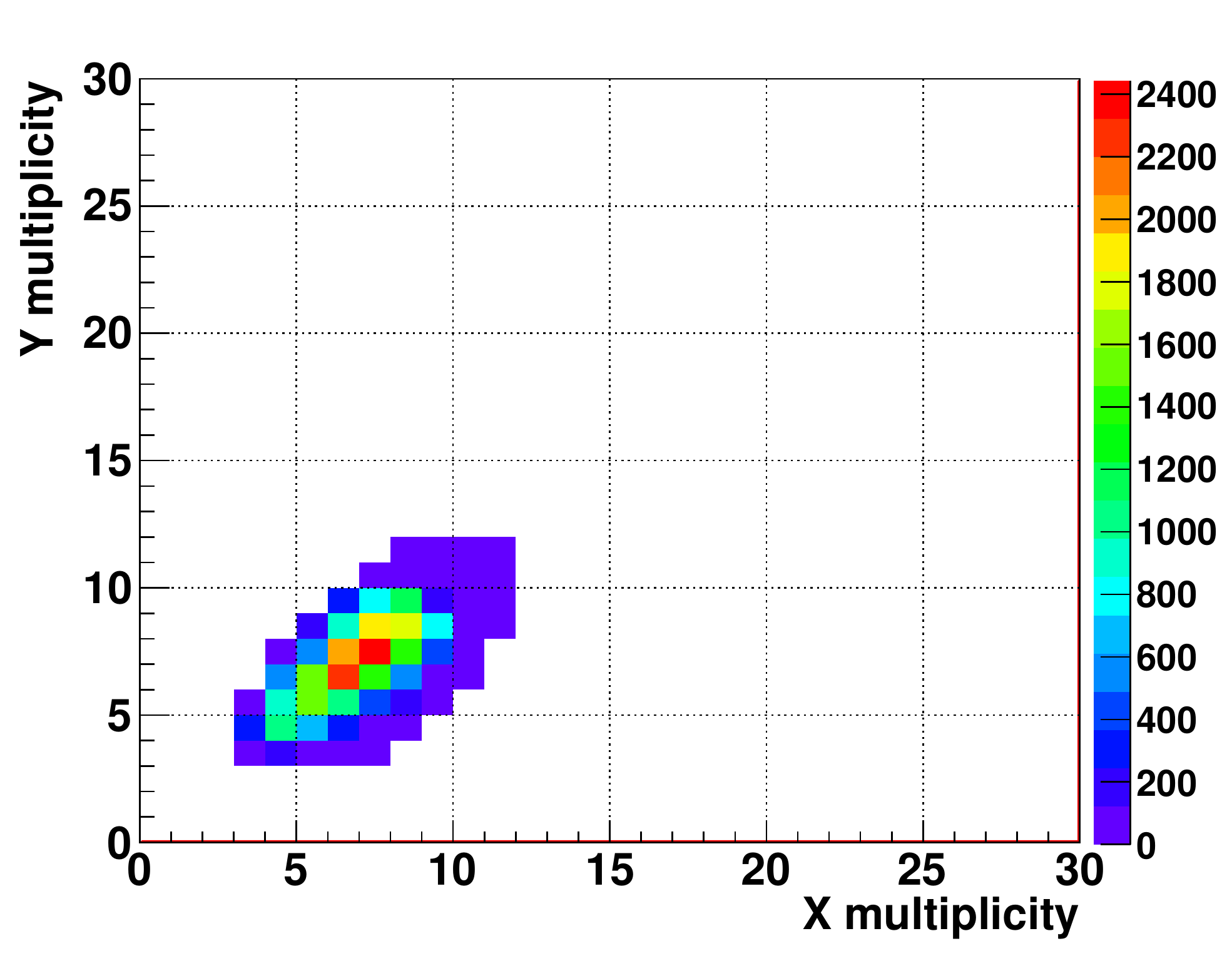}
\includegraphics[height=39mm]{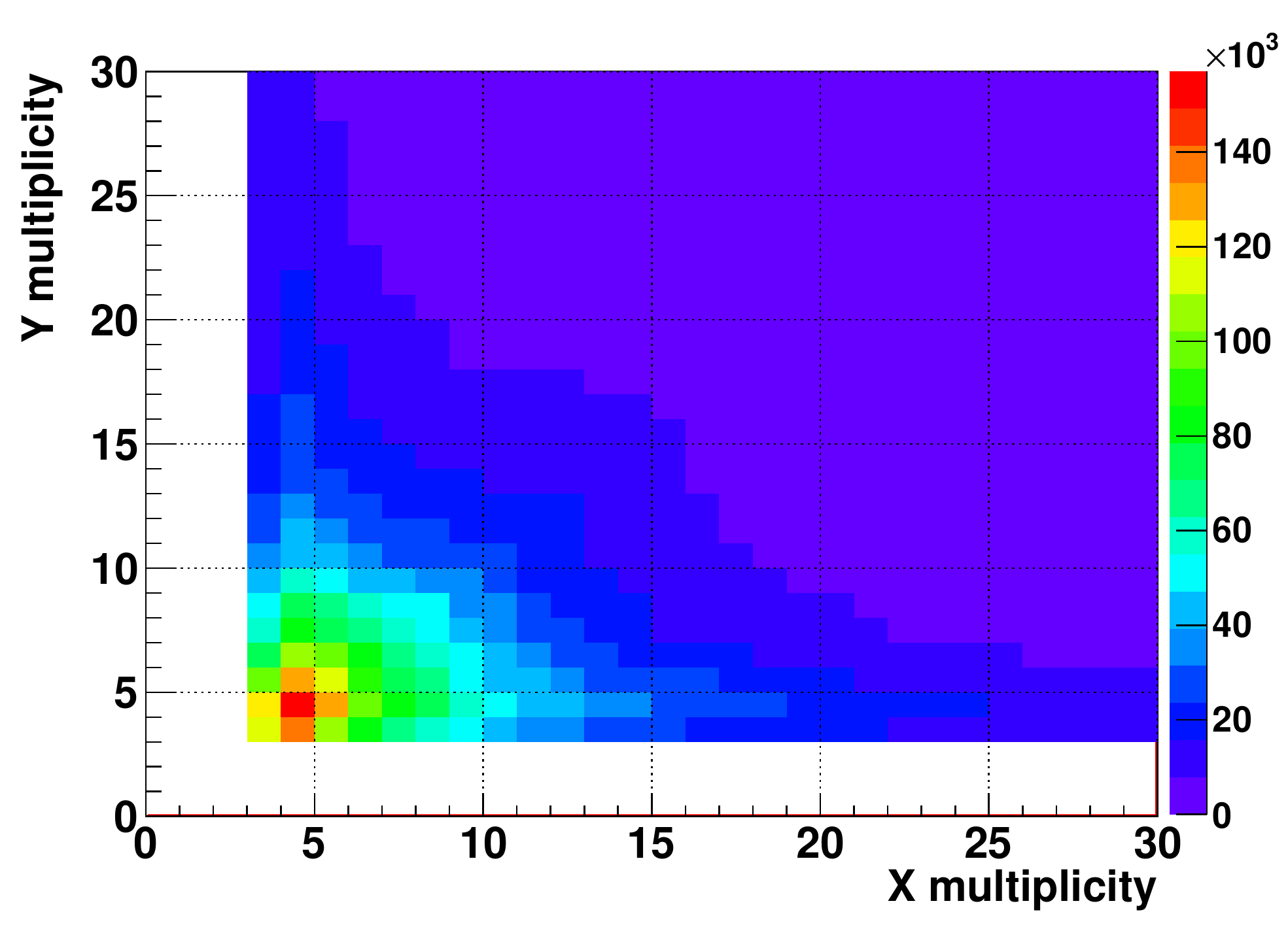}
\caption{Left: Active surface of the CAST Sunrise detector, generated with the calibration events in 2010. The horizontal and vertical lines are the images of the drift frame. No event appears at the corners due to border effects of the detector. Center and right: Number of strips activated in each direction for X-rays generated by a $^{55}$Fe calibration source (center) and by background events (right) in the CAST Sunrise detector in 2010.}
\label{fig:CASTcuts}
\end{figure}

\medskip
For each calibration run, a selection area containg the 95\% of the events is generated for the following parameters: width and number of strips activated in each cluster, for the strips; risetime, pulse width and baseline fluctuation, for the mesh pulse. These areas are used as selection criteria in background runs for rejecting cosmic muons and high energy gammas. As an example, the selection area generated by calibration for the number of strips activated in each direction and the background events' distribution is shown in figure \ref{fig:CASTcuts} (respectivelly center and right). In the same figure (left), the active area of the CAST Sunrise detector is also shown. Three other criteria are also used in the analysis: the number of clusters and the mesh pile-up (complementary to the other ones) and the baseline fluctuation (used for rejecting noisy events). 

\begin{figure}[htb!]
\centering
\includegraphics[width=77mm]{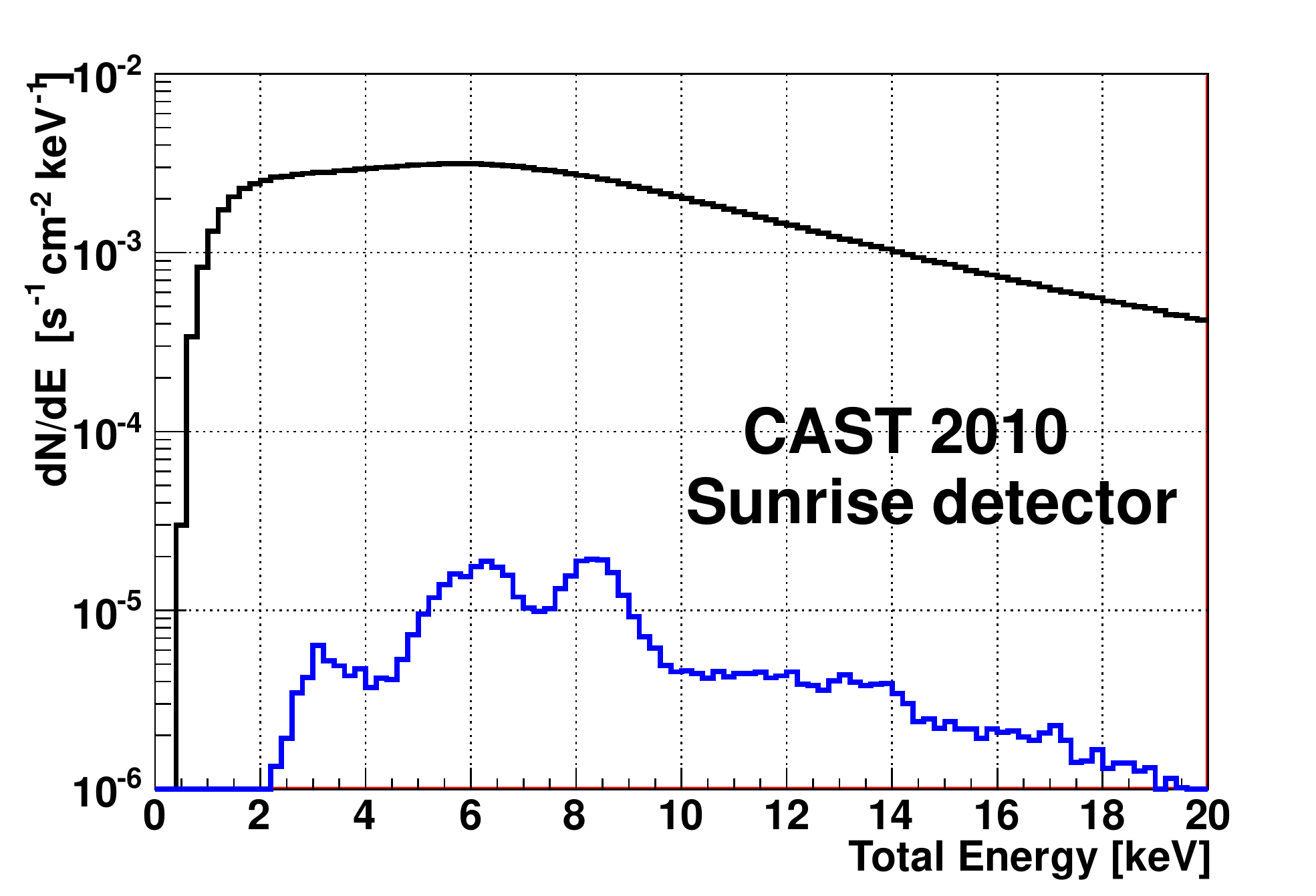}
\hspace{5mm}
\includegraphics[width=75mm]{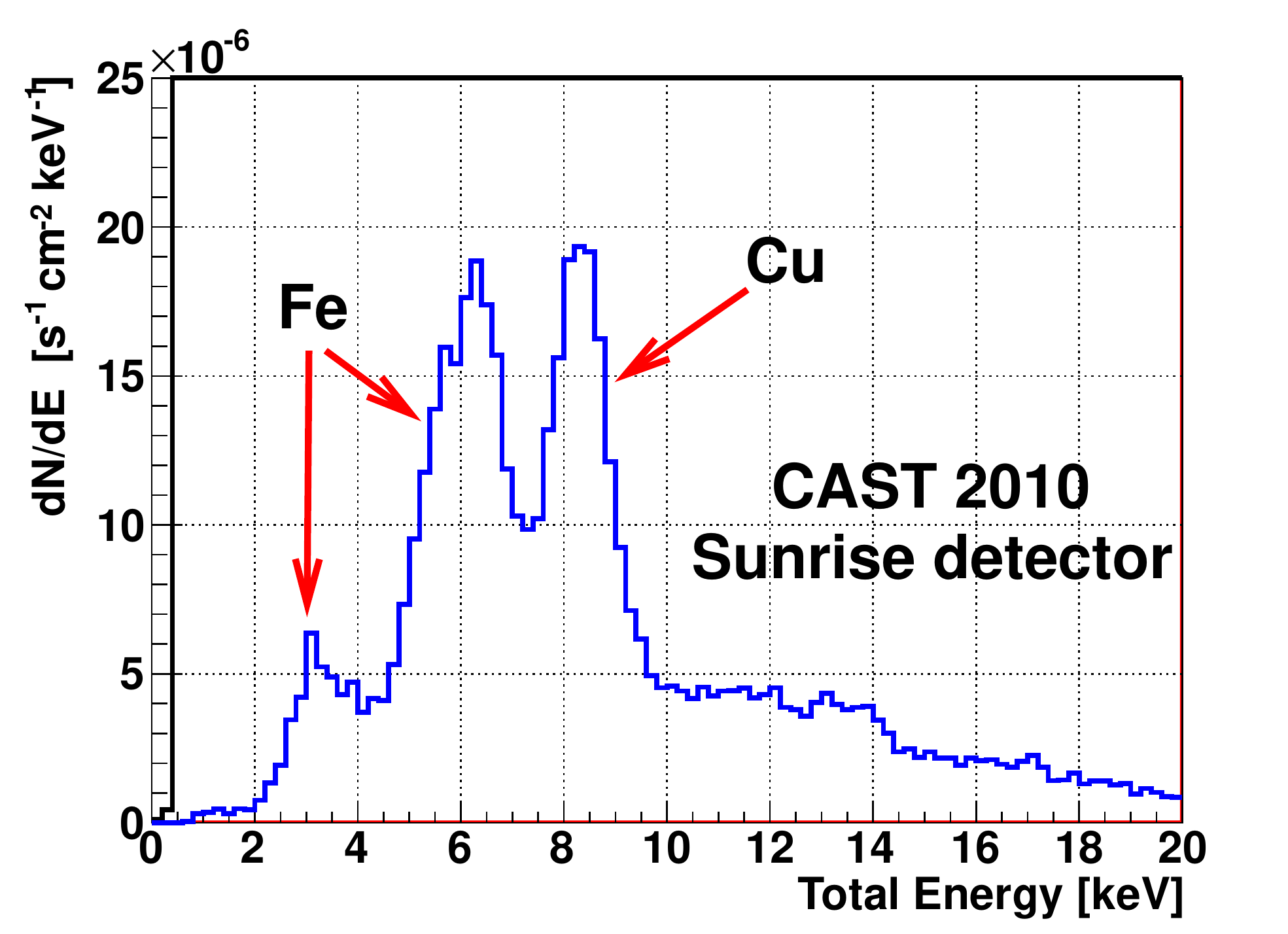}
\caption{\it Left: Background energy spectrum of the CAST Sunrise detector between 0 and 20 keV before (black line) and after the application of the selection criteria (blue line). Right: A zoom of the final background spectrum. The three peaks are generated by the fluorescence lines of the near materials: iron at 3 and 6 keV, copper at 8 keV.}
\label{CASTAnalysis}
\end{figure}

\medskip
The raw trigger rate of Micromegas detectors in CAST is around 1 Hz. Most of the background events like muons are rejected in the offline analysis with a discrimination algorithm, which keeps only events compatible with X-rays. As shown in figure \ref{CASTAnalysis} (left), the background level is reduced at least by one order of magnitude for all energies, while keeping a 80\% of 6 keV X-rays from the iron source in the calibration runs. An example of the background level in CAST is shown in figure \ref{CASTAnalysis} (right). The remaining spectrum consists mainly in three peaks at 3, 6 and 8 keV, generated by the fluorescence lines of the Micromegas' copper and the stainless steel of the window and the tube coming from the magnet bore.

\medskip
A Micromegas CAST-like detector has been installed at the Canfranc Underground Laboratory with the objective of determining the ultimate background level of this kind of readouts. A detailed description of the setup and the latest results can be found in \cite{Tomas:2011at}. Based on these results, a new Micromegas detector is being designed with improved features. The actual design supposes two main background sources in the actual detector: the radon which diffuses inside the shielding and the materials situated near the readout. The lead shielding of the actual setup will be replaced by a lead gas-tight vessel, which will contain the argon-isobutane mixture and will conserve the shielding power. The inner vessel, composed of two pieces of stainless steel and plexiglass, will be replaced by a drift cage of copper and peek. In this way, a uniform drift field will be guaranteed and there will be less and cleaner material near the readout. Finally, strips will be digitized by the T2K electronics \cite{PBaron10, PBaron09}. This new acquisition system will make the detector a real TPC, improving the background reduction.

\section{MIMAC: a dark matter experiment}
\label{sec:mimac}
The MIMAC (MIcro TPC MAtrix of Chambers) collaboration \cite{Santos:2010ds} aims at building a directional Dark Matter detector composed of a matrix of Micromegas detectors. The MIMAC project is designed to measure both 3D track and ionization energy of recoiling nuclei, thus leading to the possibility to achieve directional dark Matter detection \cite{Spergel:1988dns}. It is indeed a promising search strategy of galactic Weakly Interacting Massive Particles (WIMPs) and several projects of detector are being developed for this goal \cite{Ahlen:2010sa}. Recent studies have shown that a low exposure CF$_4$ directional detector could lead either to a competitive exclusion \cite{Billard:2010jb}, a high significance discovery \cite{Billard:2010jb2}, or even an identification of Dark Matter \cite{Billard:2011jb}, depending on the value of the WIMP-nucleon axial cross section.

\subsection{Setup description}
\begin{figure}[htb!]
\centering
\includegraphics[width=54mm]{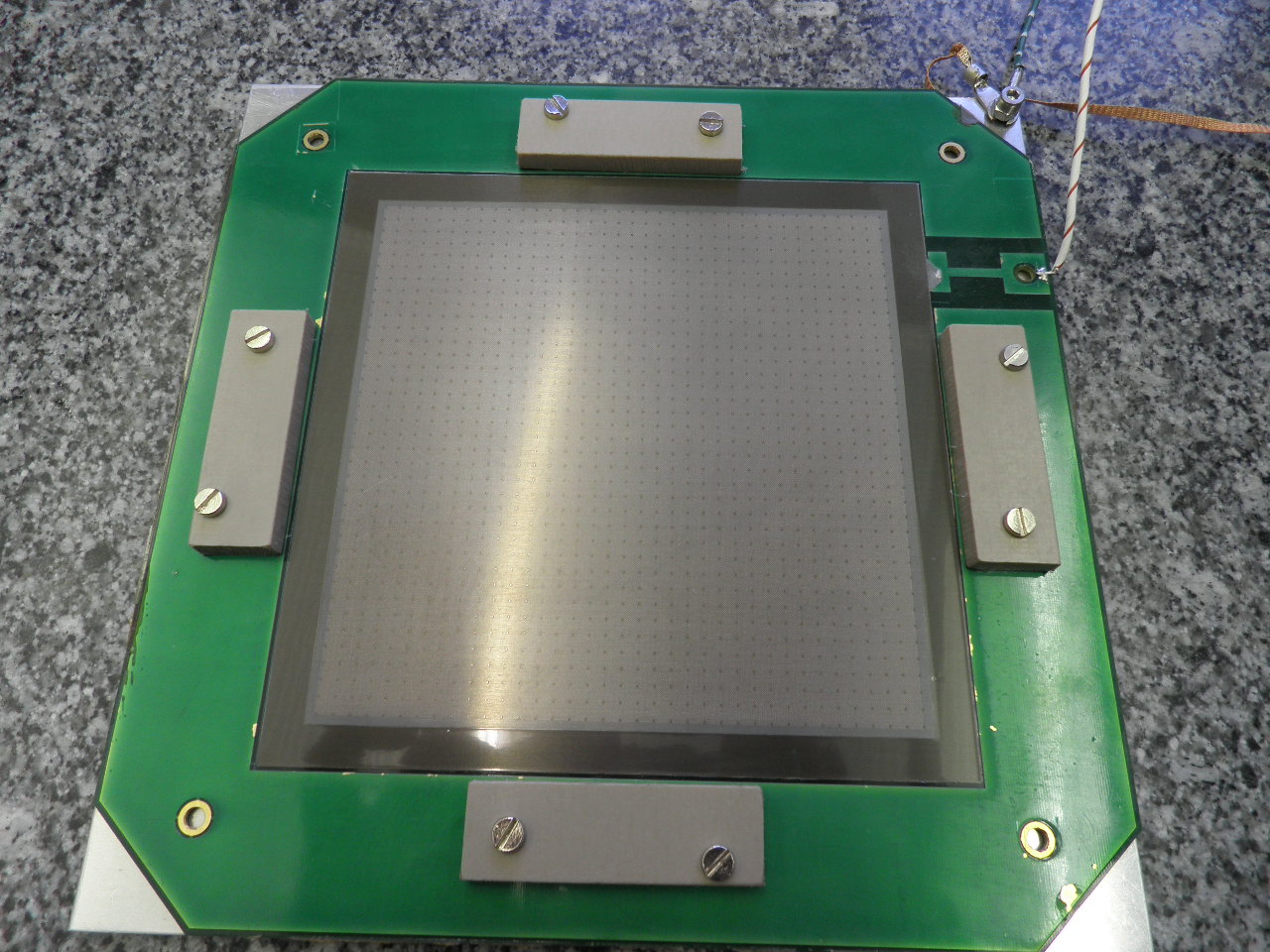}
\includegraphics[width=54mm]{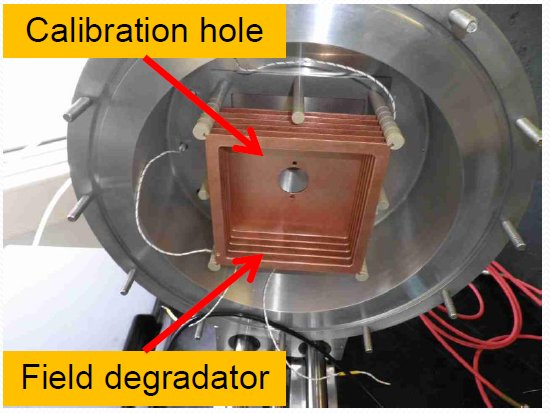}
\includegraphics[width=54mm]{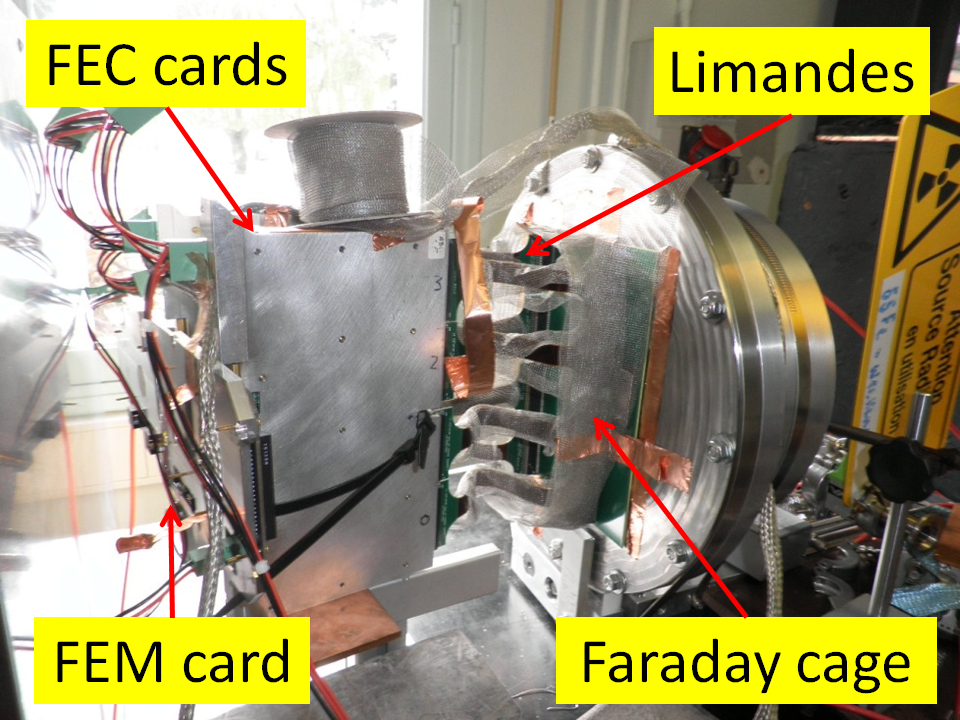}
\caption{\it Left: A view of the MIMAC bulk detector. The strips signals are rooted into 4 connectors prints at the sides of the Readout PCB, covered in the image by four plastic pieces. Center: The field degradator (made of peek bars, squared copper rings and resistors of 33 M$\Omega$). Right: A view of the dedicated vessel used to test MIMAC readouts when reading the strips with the T2K electronics. A detailed description is made in text.}
\label{fig:T2KElec}
\end{figure}

The MIMAC detector is a bulk Micromegas readout (figure \ref{fig:T2KElec}, left). It has been built from a Printed Circuit Board (PCB), whose active surface is of $10.8 \times 10.8$~cm$^2$ and has 256 strips per direction. The charge collection strips make-up an x-y structure out of electrically connected pads in the diagonal direction through metallized holes. The amplification gap is 256 $\mu$m and the strips pitch is 424 $\mu$m.The readout has been installed in a dedicated vessel, shown closed in figure \ref{fig:T2KElec} (right). It consists of two pieces which are screwed to create the TPC. One side includes an iso-KF25 valve, two gas entrances and four SHV electrical connections. At the other piece, the detector is screwed to a MIMAC bride with an interface pcb piece which keeps tight the vessel and assures the connectivity between the strips and the T2K electronics via an interface card. The vessel is equipped with a field degradator (figure \ref{fig:T2KElec}, center), which makes the drift field uniform in the conversion volume of 6 cm of height. A detailed description of the setup is made in \cite{Iguaz:2011fi2}. The detector was characterized in Ar+5\%iC$_4$H$_10$, showing an energy resolution of 18\% FWHM at 5.9 keV.

\begin{figure}[htb!]
\centering
\includegraphics[width=75mm]{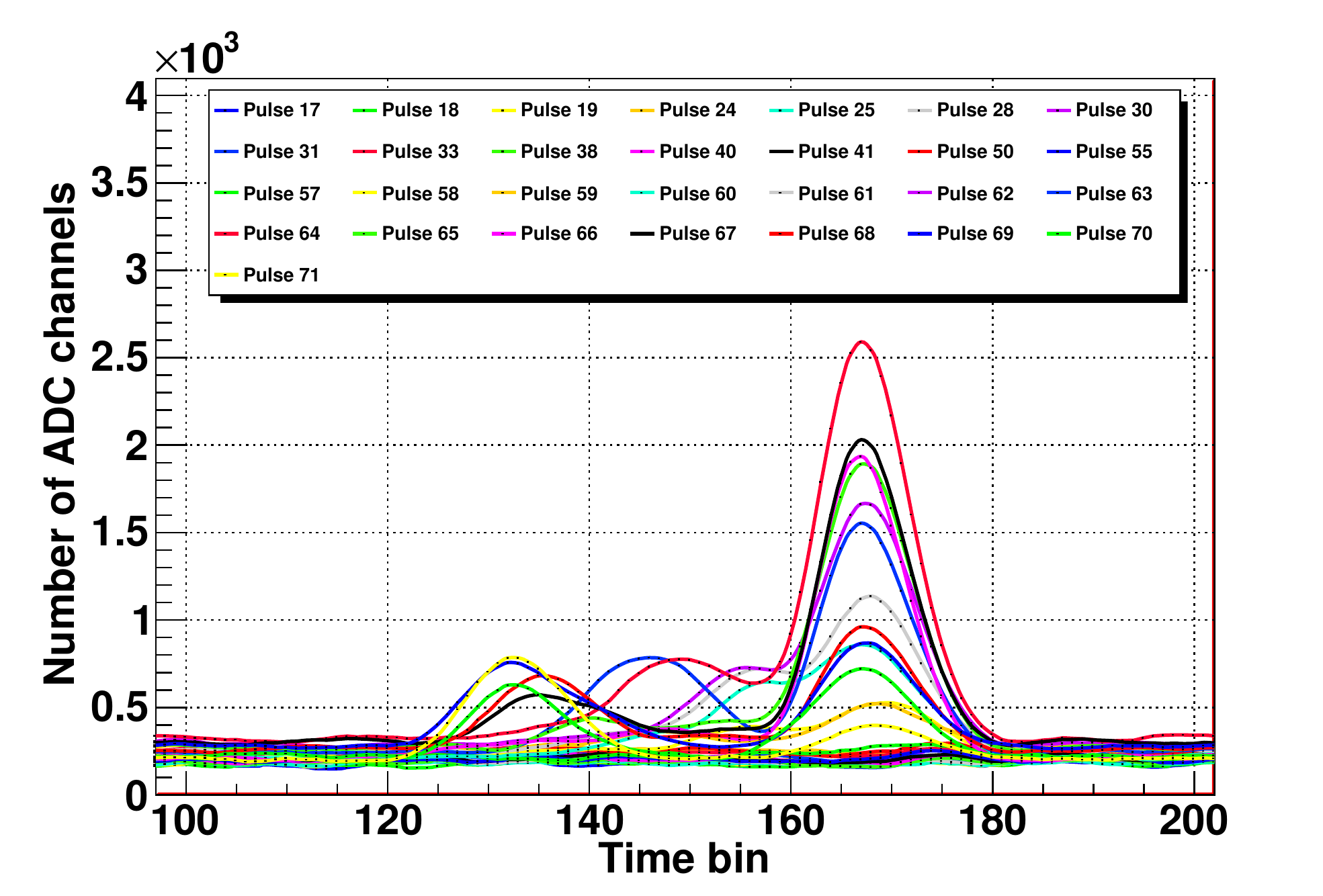}
\hspace{10mm}
\includegraphics[width=75mm]{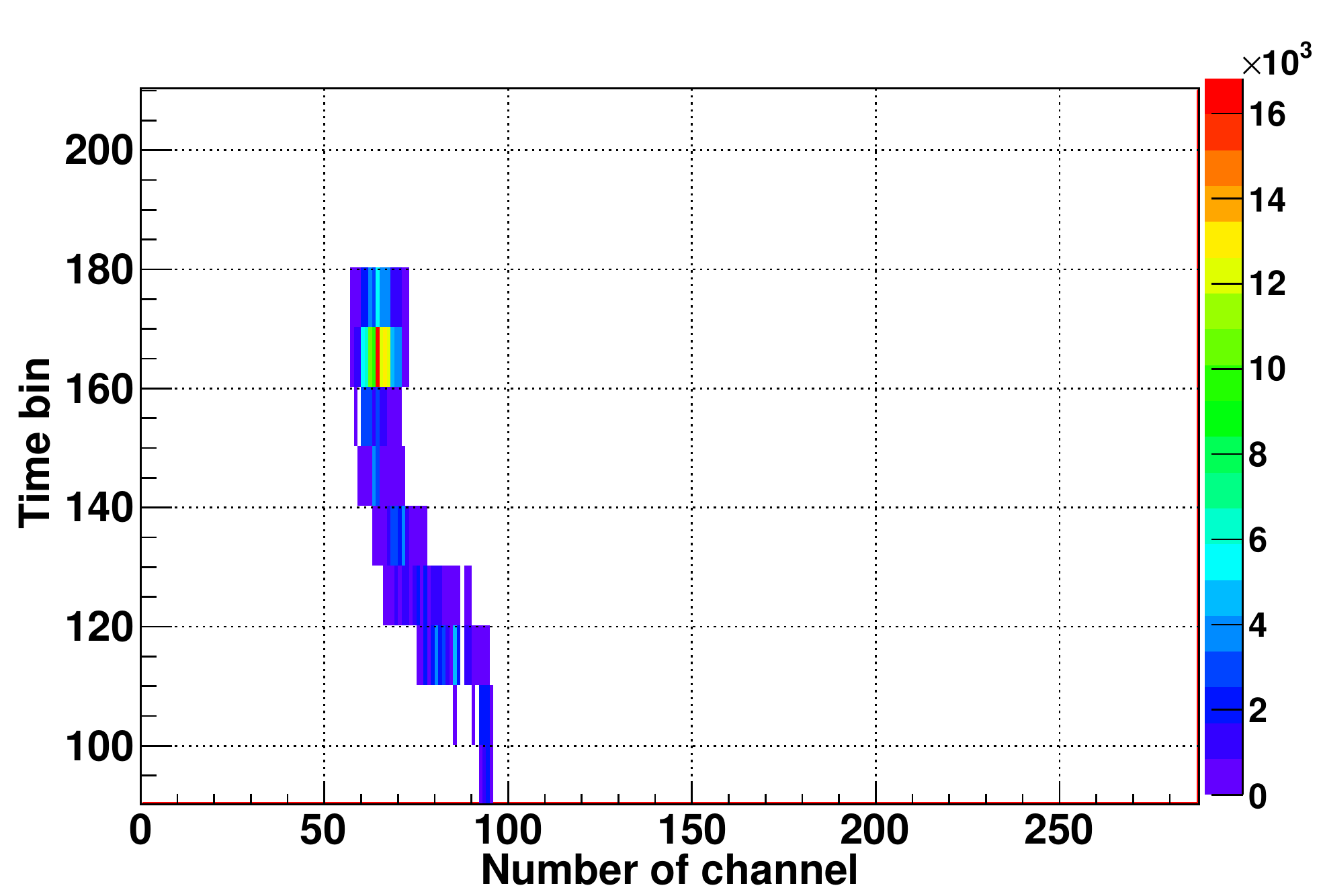}
\caption{\it Left: Example of pulses induced in the strips, acquired with the T2K electronics. Right: The reconstruction of the XZ projection of the same event. The physical event is an electron of some keV with a final charge accumulation (or blob).}
\label{fig:PulseEvent}
\end{figure}

\medskip
The T2K electronics \cite{PBaron10, PBaron09} has been used to read the the signals induced in the strips and to fully validate the concept of MIMAC readouts. Eight cables (so-called {\it limandes}) take the strips signals into two FEC cards. Each card contains four ASIC chips which digitize in 511 samples the signals of 72 channels, which are previously amplified and shaped. Finally, the data of each ASIC is sent by a FEM card to a DAQ card and subsequently to the computer for recording. As the external trigger mode of the T2K DAQ has been used, a trigger signal has been created feeding the bipolar output of the ORTEC 472A Spectroscopy amplifier (which is used to amplify the signal induced at the mesh) into a FAN IN/OUT Lecroy 428F and subsequently into a NIM-TTL converter. Strips pulses have been sampled every 20 ns and the shapping time has been fixed to 100 ns. An offline analysis programme has been developped to extract the strips pulses from the raw files generated by the DAQ and to record them into a ROOT-like file \cite{ROOT:1996}. The same programme reconstructs the two 2D projection of each event from the strips pulses, using the amplitude of each pulse sample and the readout decoding. An example of the strips pulses and the XZ reconstruction of one event is shown in figure \ref{fig:PulseEvent}.

\subsection{Discrimination of low energy x-rays}
\medskip
One of the main objectives of Dark Matter experiments is the reduction of the background level using any kind of selection criteria. TPCs and micromegas readouts in particular have the advantage that the geometry of each event can be observed. In our tests, we have used an argon based mixture, which makes quite improbable to detect nuclear recoils of more than 100 keV. Meanwhile, neutrons of less than 100 keV produce tracks of maximum 300 $\mu$m length at atmospheric pressure, which is less than the pitch of our readout. For these reasons, we have focused on discriminating low-energy photons from the rest of backgrounds events like muons, alphas and high energy gammas.

\medskip
During three weeks of data-taking, a constant flow of 5 l/h of Ar+5\%iC$_4$H$_{10}$ circulated by the dedicated vessel. The detector of the 256 $\mu$m-gap was kept in tension ($E_{amp} = 21.9 kV/cm$, $E_{drift} = 88 V/cm$) and acquiring background events in a continous way. The readout was calibrated twice per day to check the evolution of the gain, energy resolution and the calculated parameters. The gain fluctuated around 10\%, due to the variations of pressure and temperature inside the vessel, as no effective control of these parameters was made. The energy resolution remained stable between 18 and 20\% FWHM during the same period.

\medskip
In each spatial direction of the two event projections, the following parameters were calculated: charge, mean position, width and number of strips activated. This analysis was then extended to the perpendicular direction using the amplitudes of strips pulse in each temporal bin. Finally, the total charge of each event was obtained summing the charge of both projections. This analysis was motivated by the CAST experiment \cite{Papaevangelou:2008tp, FJIguaz:2010}, which looks for axion-induced photons of less than 10 keV. However, the CAST electronics has no temporal information in each strip and can only use instead the features of the mesh induced pulses for time information.

\begin{figure}[htb!]
\centering
\includegraphics[width=75mm]{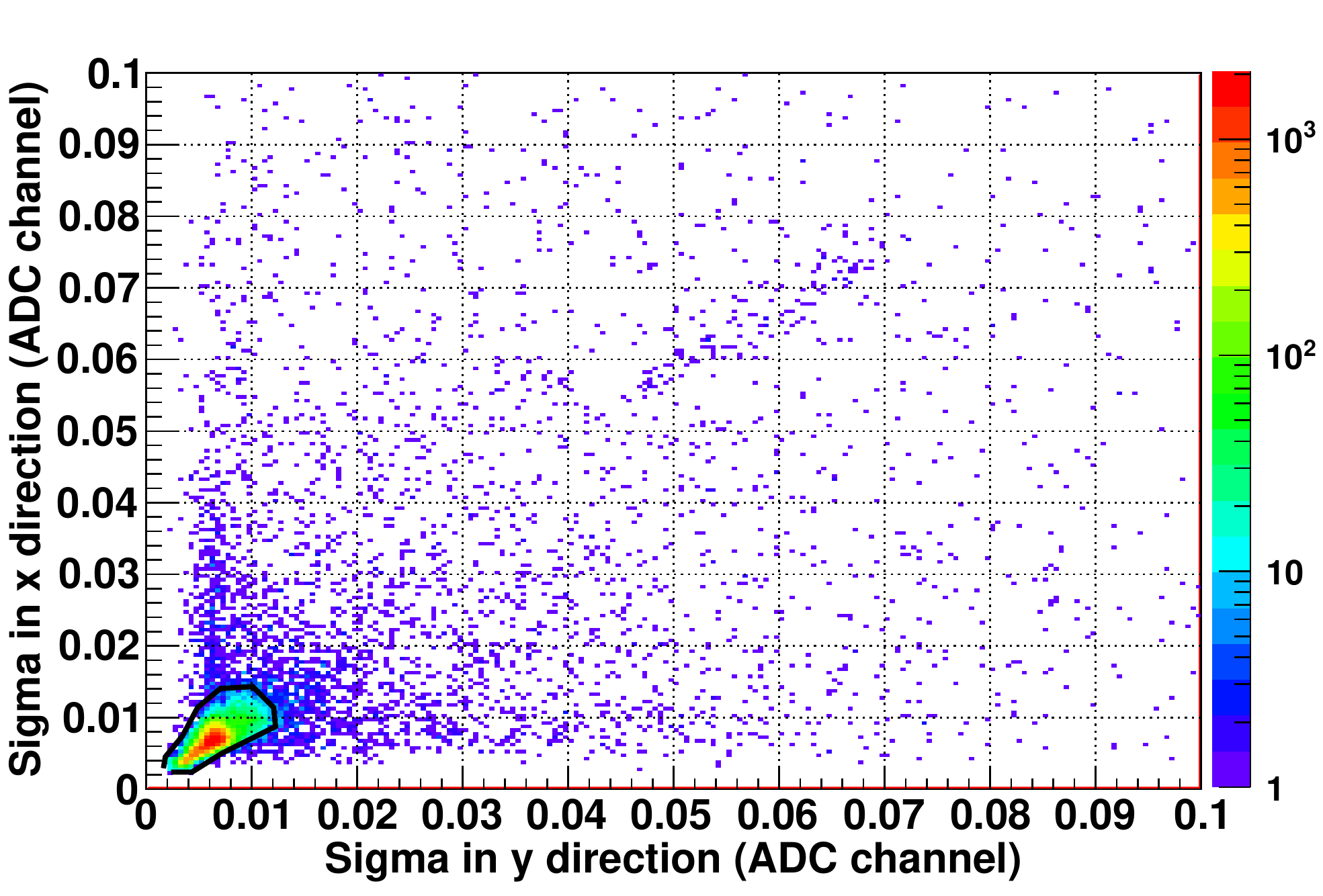}
\hspace{10mm}
\includegraphics[width=75mm]{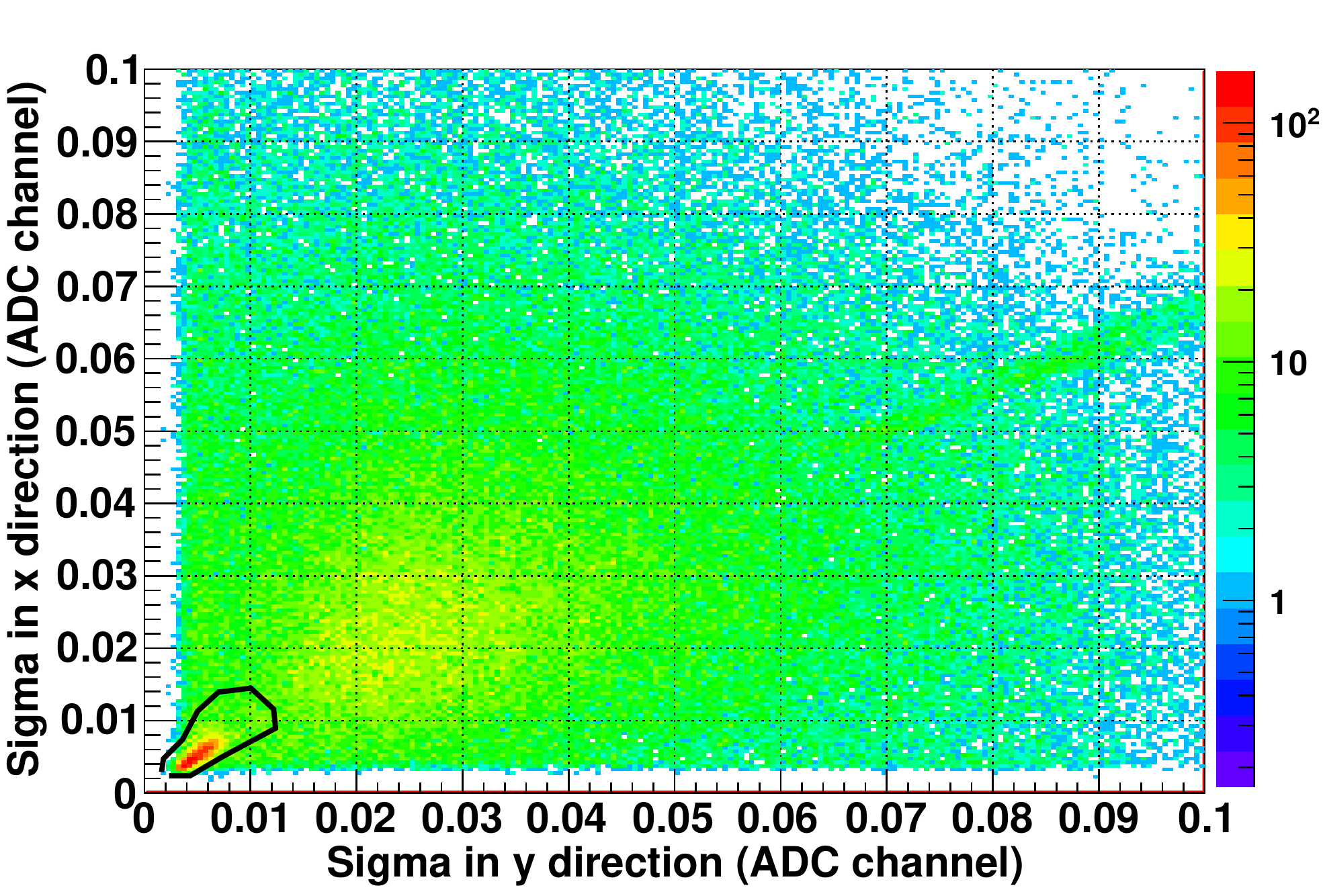}
\caption{\it 2D distribution of the spatial width in X and Y directions for calibration (left) and background events (right). The thick black line limits the region used for discriminating low energy photons from the rest of background events.}
\label{fig:SigmaXY}
\end{figure}

\medskip
An x-ray of less than 10 keV produces a primary ionization localized in a range of less than 1 mm. The amplification of this signals creates a distribution of charge (called {\it clusters}), with a short width and few strips activated. In contrast, cosmic muons and high energy gammas produce a spatially extended ionization, resulting in broader cluster and more strips activated. As an example of these differences, the cluster width in both spatial directions for $^{55}$Fe photons and background events is shown in figure \ref{fig:SigmaXY}. Using both distributions, a selection area was defined to reject the maximum number of background events and with the lowest rejection of low energy photons. The same comparison was made with the distributions of the number of strips activated and the number of temporal samples, resulting in two other selection areas.

\begin{figure}[htb!]
\centering
\includegraphics[width=75mm]{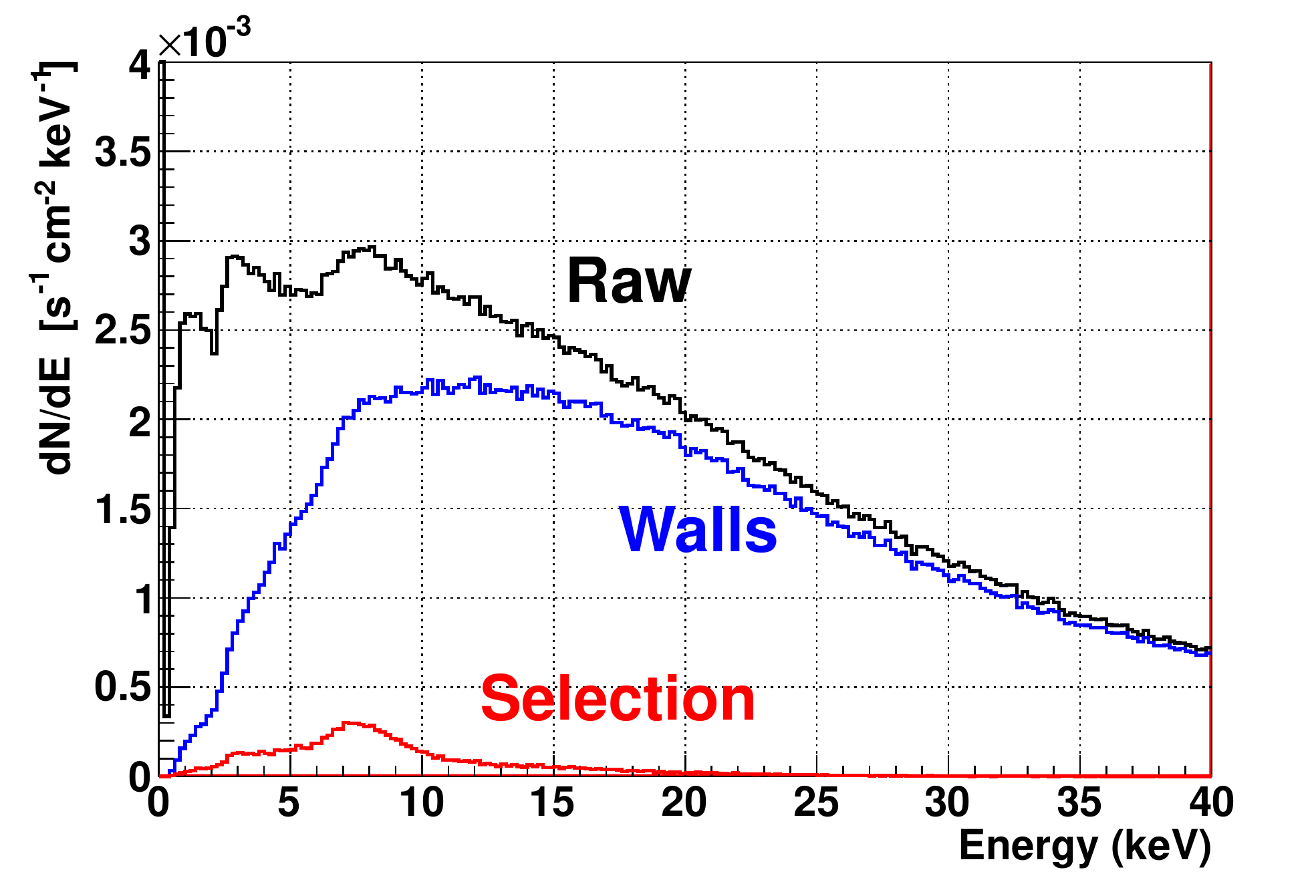}
\hspace{10mm}
\includegraphics[width=75mm]{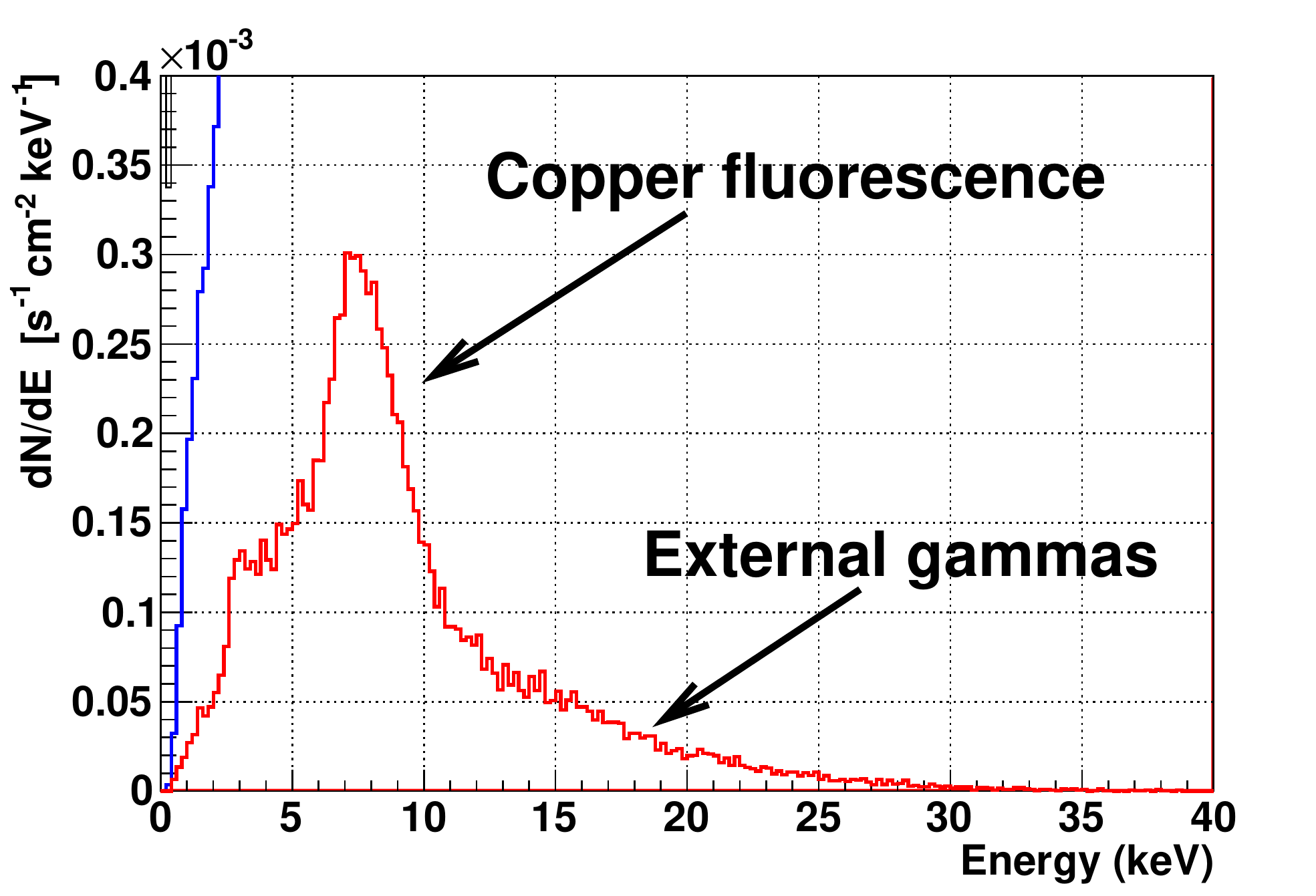}
\caption{\it Left: Background energy spectrum between 0 and 40 keV before (black line) and after the succesive application of the walls (blue line) and the strips criteria (red line). Right: A zoom of the final background spectrum, generated mainly by the copper fluorescence line at 8 keV and the environmental gammas.}
\label{fig:Spectrum10x10}
\end{figure}

\medskip
Apart from these three selection criteria, events near the borders of the detector or with less than 2 strips activated in any direction (generally noisy events) were vetoed. The rejection factor of these cuts is shown in figure \ref{fig:Spectrum10x10}. The difference between the spectra before and after the application of these criteria is respectively a factor 20 and 12 for 2-6 keV and 6-10 keV. If only the strips criteria are considered, this value is a factor 8.6 for both energy ranges. Meanwhile, the signal efficiency is respectively 78\% and 86.7\% at 3 and 6 keV, a bit better than CAST values. The remaining background spectrum, shown at the same figure, is dominated by the environmental gammas and the copper fluorescence line at 8 keV, activated by muons and gammas hitting the micromegas readout.

\section{Neutron and gamma discrimination}
Micromegas detectors have a high granularity, which allows them to register the event topology in good detail. This feature can be used to discriminate signals from background events with high efficiency. In a feasibility study of Dark Matter TPC with Micromegas, the first step that must be proved is that gamma and nuclear recoils can be easily discriminated. This topic was firstly studied in \cite{Tomas:2007at}. We must note that the charge distribution created by low energy gammas at low gas pressure is not anymore point-like but asymmetric. As a consequence, these events are wider in one direction than the others and the maximum width can be used as a discrimination parameter.

\begin{figure}[htb!]
\centering
\includegraphics[width=75mm]{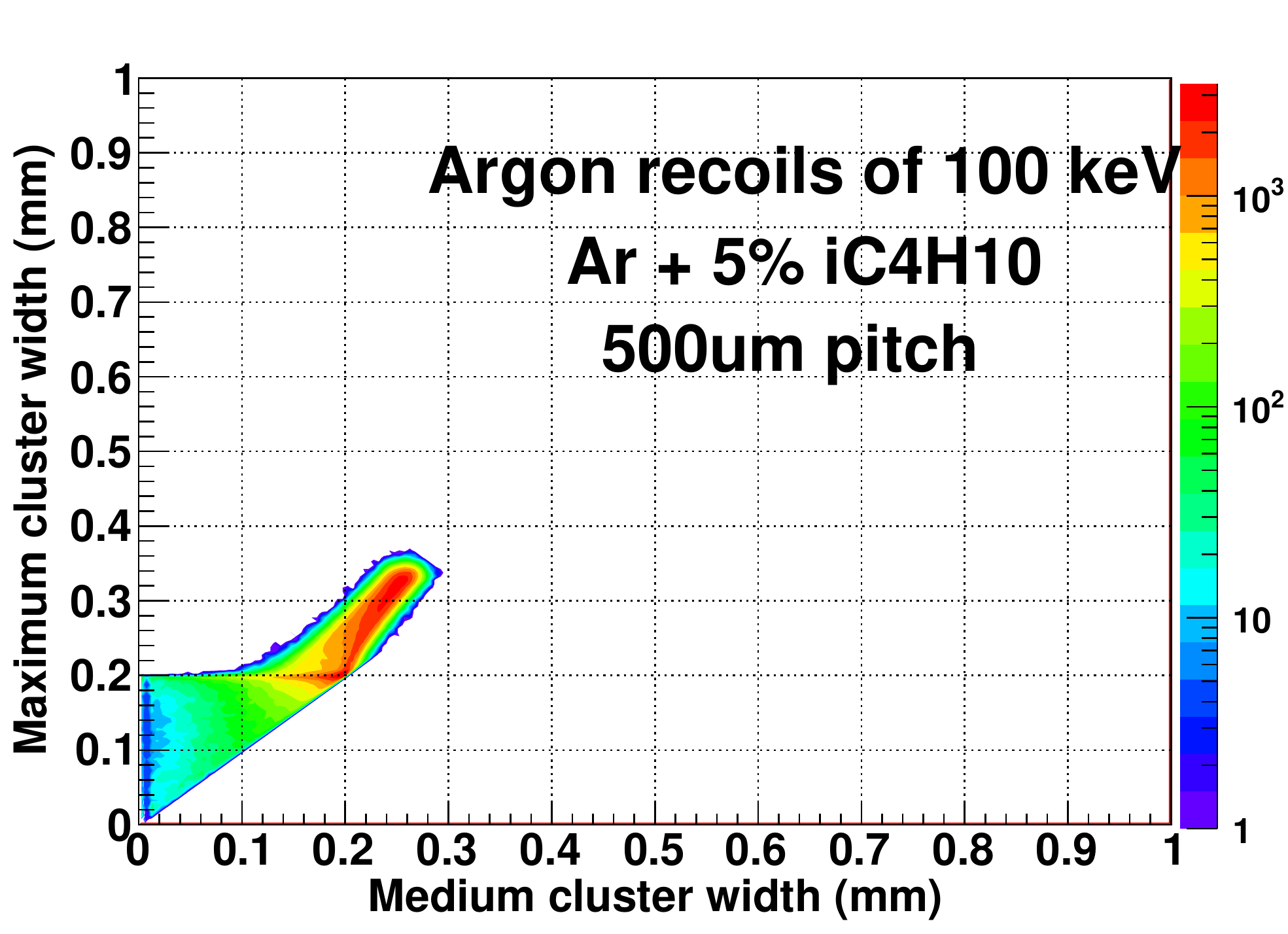}
\hspace{5mm}
\includegraphics[width=75mm]{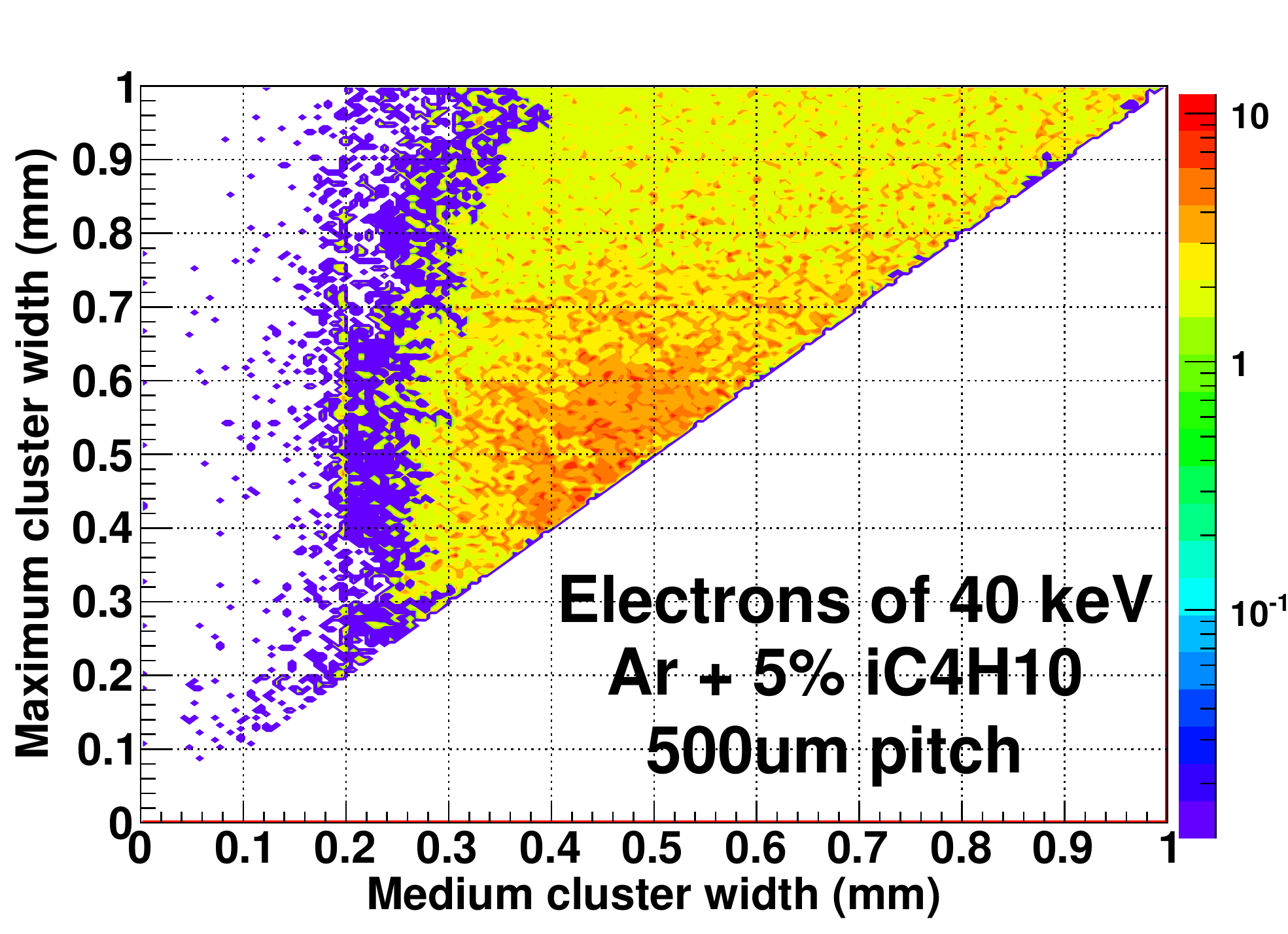}
\caption{\it Dependence of the maximum with the medium cluster width for argon recoils of 100 keV (left) and electrons of 40 keV (right) in a TPC filled with Ar+5\%iC$_4$H$_{10}$ at atmospheric pressure and equipped with a microbulk detector of 500 $\mu$m pitch.}
\label{fig:NeuGamma}
\end{figure}

\begin{figure}[htb!]
\centering
\includegraphics[width=75mm]{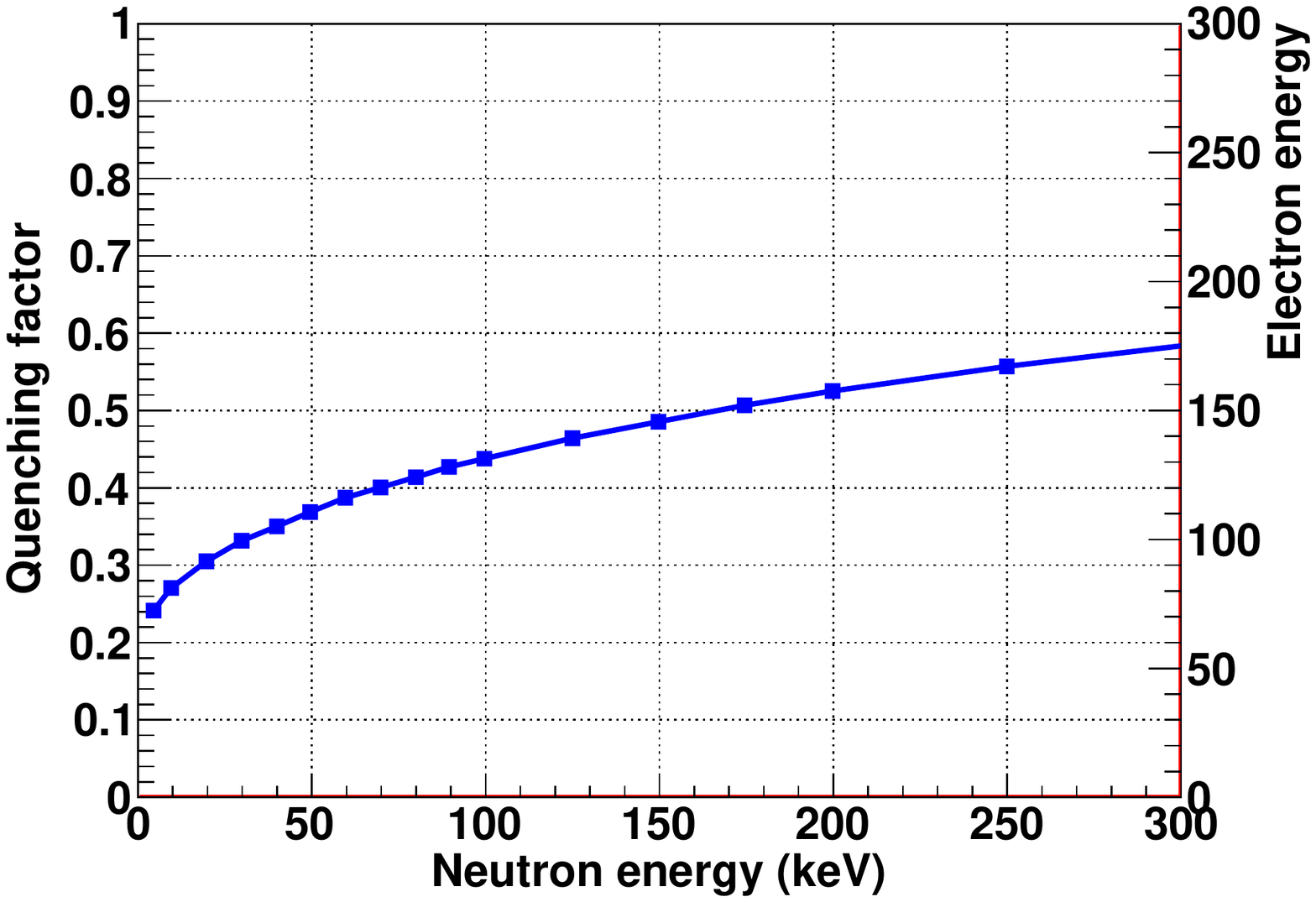}
\hspace{5mm}
\includegraphics[width=75mm]{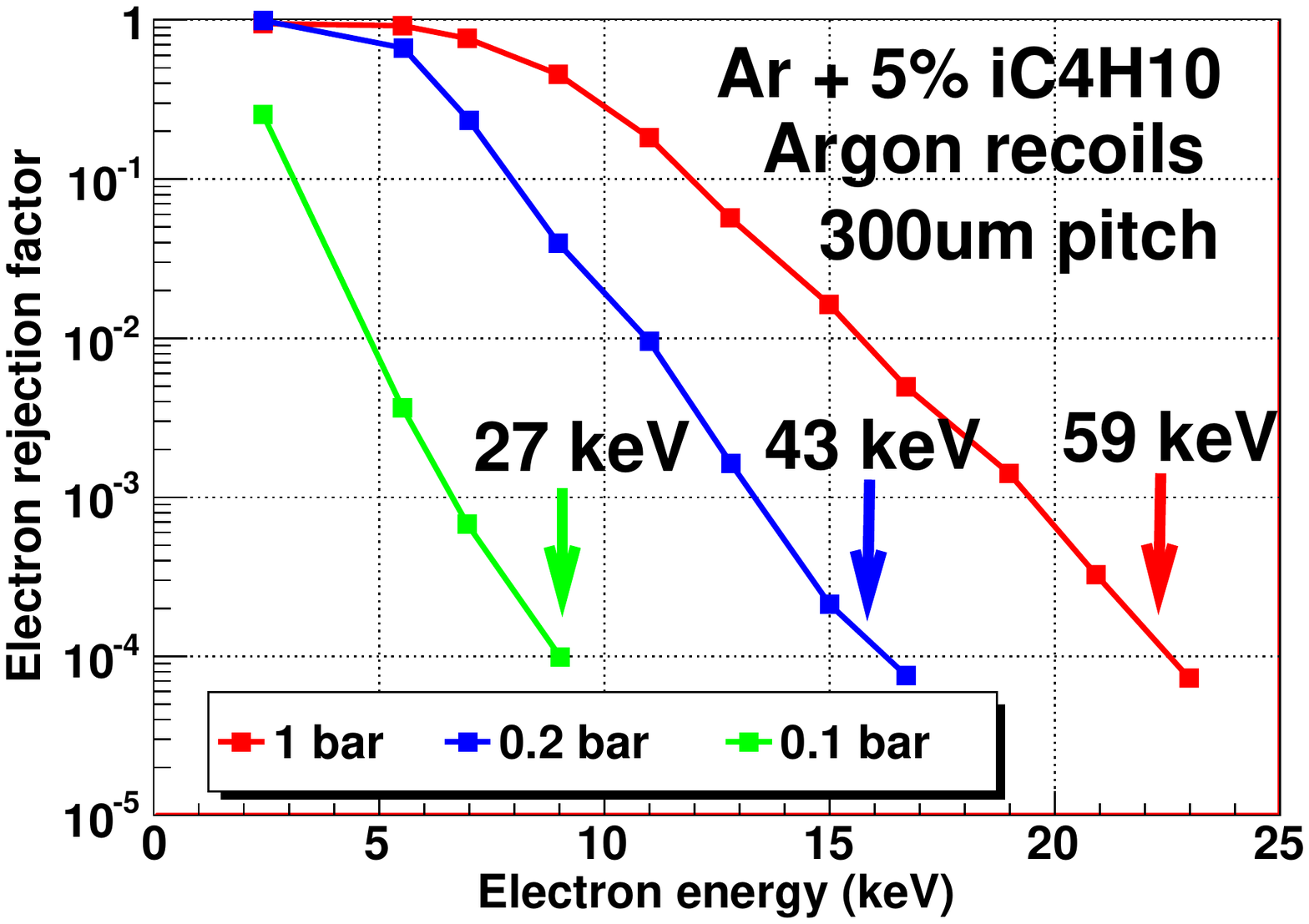}
\caption{\it Left: Dependence of the equivalent electron energy and the quenching factor with the neutron energy for pure argon, calculated with Lindhard's theory \cite{Lindhard:2010jl}. Right: Dependence of the electron rejection factor with the electron energy in a TPC filled with Ar+5\%iC$_4$H$_{10}$ and equipped with a microbulk detector of 500$\mu$m pitch for different gas pressures.}
\label{fig:Rej}
\end{figure}

\medskip
Argon recoils with energies between 10 and 100 keV and electrons between 2 and 40 keV were simulated in a TPC equipped with a microbulk readout, with an amplification gap of 50 $\mu$m and a strips pitch of 500 $\mu$m. The TPC was a cubic box of 20 cm-length and was filled with Ar+5\%iC$_4$H$_{10}$ and pressures between 0.2 and 1 bar. A detailed description of the simulation is made in \cite{Iguaz:2010fi}. For each neutron energy and pressure, the dependence of the maximum and the medium cluster widths showed a limited area, as shown in figure \ref{fig:NeuGamma} (left). In contrast, the electrons of equivalent energy, calculated by Lindhard's theory (shown in figure \ref{fig:Rej}, left), showed a distribution that clearly separated from the former one. Both distributions approach to each other at low energies because electron events are more symmetric and their width is basically defined by the diffusion.

\medskip
For each gas pressure and neutron energy, the selection areas defined by argon recoils were delimited. The number of electron events whose parameters were inside these areas was then calculated and the electron rejection factor was calculated. The dependence of this parameter with the electron energy and the gas pressure is shown in figure \ref{fig:Rej} (right). For each pressure, the higher energy electrons have, the more efficient they are rejected. This factor increases if the gas pressure is decreased. For example, if a minimum rejection factor of $10^{-4}$ is considered, the electron energy threshold is respectively at 9, 16 and 22 keV at 0.2, 0.5 and 1 bar, which corresponds to a neutron energy threshold of 27, 43 and 59 keV. These values are better for lighter gases like CF$_4$ or Ne.

\section{Conclusions}
\label{sec:conc}
Micromegas detectors are being used in Dark Matter Experiments due to their good discrimination capabilites for low energy events. In the CAST experiment, three microbulk detectors are installed since 2007. Muons are efficiently rejected by the offline analysis and the background is mainly due to the fluorescence lines of near materials. A CAST detector has been installed in the LSC to find its ultimate background level. A new design is being developped based on these results, with several improvements.

\medskip
Another application of Micromegas technology in this field is MIMAC project. It aims at building a directional Dark Matter detector composed of a matrix of Micromegas detectors. These readouts will measure both 3D track and ionization energy of recoiling nuclei, thus leading to the possibility to achieve directional dark Matter detection. A $10\times 10$ cm$^2$ readout has been completely validated with the T2K electronics. Low energy photons (2-10 keV) have been selected rejecting events with larger topologies. In the near future, the capability of Micromegas detectors to select neutrons frow low energy electrons at low pressure will be studied.

\section*{Acknowledges}
\label{sec:ack}
We acknowledge the ANR-07-BLANC-0255-03 funding. The authors would like to thank D. Desforges for his availability in the use of the Mitutuyo Microscope as well as D. Jourde for his help with the degrador.


\end{document}